\documentstyle[aps,prl,preprint]{revtex}

\draft

\begin{document}

\title{Do Spectral Trace Formul\ae\ Converge?}
\author{Eyal Doron}
\address{Max-Planck Institut f\"ur Kernphysik, Postfach 103980,
69029 Heidelberg, F.R.G.}               

\date{\today}
\maketitle

\begin{abstract}

We evaluate the Gutzwiller trace formula for the level density of
classically chaotic systems by considering the level density in a bounded
energy range and truncating its Fourier integral. This results in a limiting
procedure which comprises a convergent semiclassical approximation to a well
defined spectral quantity at each stage. We test this result on the spectrum
of zeros of the Riemann zeta function, obtaining increasingly good
approximations to the level density. The Fourier approach also explains the
origin of the convergence problems encountered by the orbit truncation
scheme.

\end{abstract}


\narrowtext

The subject of ``quantum chaos" comprises the study of quantum systems whose
classical analogue is chaotic. One of the main tools is the Gutzwiller trace
formula, which provides a semiclassical approximation (SCA) to the
fluctuating part of the quantum level density via an infinite sum over
classical periodic orbits. A major problem with this sum is that it does not
converge absolutely, making manipulations of the sum rather suspect, and
moreover dependent on the ordering.

An important question in this context is whether the Gutzwiller sum can be
made to {\em conditionally} converge, by using an appropriate truncation
procedure and possibly a reordering. Convergence in this sense means that
one obtains better and better approximations to the level density, limited
only by the SCA itself, as the truncation point goes to infinity. The most
natural truncation scheme is to include only orbits whose period is bounded
by some $T_{\text{max}}$ (we will call this scheme ``orbit truncation").
Unfortunately, this has not had much numerical success.  Gutzwiller in
\cite{Gutzwiller88} used orbit truncation to evaluate the semiclassical
level density for the anisotropic Kepler problem, and was able to reproduce
the level density for low energies but not for higher ones --- an
incongruous result for an SCA.  A mathematical model which is analogous to
chaotic systems, the zeros of Riemann's zeta function (see below), displays
similar behavior: as one adds ``periodic orbits" (prime numbers, in this
case) to the sum, peaks in the level density become submerged in increasing
oscillations \cite{KeatingVarenna}. In the hyperbola billiard there are
indications that the sum as written does conditionally converge, however
this is an exceptional system in which the Lyapunov exponent is larger than
the Kolmogorov entropy \cite{MartinPhD}.

In this work we treat the problem of conditional convergence from a slightly
different point of view. We consider $\tilde d(\varepsilon)$, the
fluctuating part of the level density in a {\em finite} energy range, and
define a purely quantum spectral quantity $\tilde d_{{\cal T}}(\varepsilon)$
which converges to $\tilde d(\varepsilon)$ as ${\cal T}\to\infty$.  $\tilde
d_{{\cal T}}(\varepsilon)$ is constructed via a smooth truncation of the
Fourier transform of $\tilde d(\varepsilon)$. We then use the Gutzwiller sum
to obtain the SCA to $\tilde d_{{\cal T}}(\varepsilon)$, which for any
${\cal T}$ involves an absolutely converging sum over periodic orbits. The
limit ${\cal T}\to\infty$ can now be taken safely, avoiding the convergence
problems encountered by orbit truncation --- the only remaining errors are
due to the SCA itself. This approach differs from previously suggested ones
\cite{Guinand48,BerrySchool,Aurich92} in that it does not involve a
resummation of the divergent tail of the sum or invoke the bootstrapping
mechanism, but rather retains the intuitively appealing ordering by orbit
period.

We test this approach ,which we call ``Fourier truncation", on the spectrum
of zeros of Riemann's zeta function, and obtain very good numerical results.
Moreover, we use the Fourier approach to analyze the orbit truncation scheme
as well, and show that it leads to incorrect evaluation of the last few
``frequency" components of the level density. Because of the exponential
proliferation of orbits, these errors can become very large and overwhelm
the resulting approximation.

Our starting point is the fluctuating level density
$\tilde d(E) = \sum_m \delta(E-E_m)-\bar d(E)$, where the $\{E_m\}$ are the
set of eigenvalues of the Hamiltonian and $\bar d(E)$ is the smoothed level
density (as given by e.g.\ Weyl's form). We will confine ourselves to totally
hyperbolic time-independent systems with a bounded phase space,
in which case the $E_m$ are discrete and correspond to eigenenergies.

We now select a finite energy range $[E_0\pm\Delta E/2]$ by means of a real
and symmetric {\em window function} $\chi(\varepsilon)$, which satisfies the
requirement $\chi(|\varepsilon|>\Delta E/2) \approx 0$. A convenient form is
the Gaussian, $\chi(\varepsilon) = \exp(-\varepsilon^2/2\sigma^2)$.  We then
define the {\em local} fluctuating part of the level density $\tilde
d(\varepsilon) = \tilde d(E_0+\varepsilon)\,\chi(\varepsilon)$, and expand
it in a Fourier integral,
\begin{mathletters}
\begin{eqnarray}
  \tilde d(\varepsilon) &=& {1\over 2\pi\hbar} \int_{-\infty}^\infty d\tau\;
     \hat d(\tau)\exp\left(-i{\tau\varepsilon\over\hbar}\right)\;,  
     \label{Deps} \\
  \hat d(\tau) &=&  \int_{-\infty}^{\infty} d\varepsilon\;
           \tilde d(\varepsilon)\exp\left(i{\tau\varepsilon\over\hbar}\right)
            \;.  \label{dtau}
\end{eqnarray}
\end{mathletters}
The value of $\hat d(\tau)$ using the true quantum level density is
\begin{equation}
  \hat d(\tau) = \sum_m \chi(\varepsilon_m)
              \exp\left(i{\tau\varepsilon_m\over\hbar}\right) \;,
    \label{ExactCoeffs}
\end{equation}
where $\varepsilon_m=E_m-E_0$.  Let us now define $\tilde d_{\cal
T}(\varepsilon)$ as the approximation to $\tilde d(\varepsilon)$ obtained by
truncating $\hat d(\tau)$ at $\tau={\cal T}$. We perform this truncation by
means of another real and symmetric window function $W_{\cal T}(\tau)$,
which satisfies $W_{\cal T}(\tau)=0$ for $|\tau|>{\cal T}$. Thus $\tilde
d(\varepsilon) = \lim_{{\cal T}\to\infty} \tilde d_{\cal T}(\varepsilon)$,
where
\begin{equation}
  \tilde d_{\cal T}(\varepsilon) = {1\over 2\pi\hbar} \int_{-\infty}^{\infty}
            d\tau\; W_{\cal T}(\tau)\,\hat d(\tau)
                \exp\left(-i{\tau\varepsilon\over\hbar}\right)\;.
                \label{PartialSum}
\end{equation}
From (\ref{ExactCoeffs}) $\tilde d_{\cal T}(\varepsilon)$ is given by
\begin{equation}
  \tilde d_{\cal T}(\varepsilon) = \sum_m \chi(\varepsilon_m)\,
       w_{\cal T}(\varepsilon_m-\varepsilon) \;,  \label{ApproxDeltas}
\end{equation}
where $w_{\cal T}(\varepsilon)=(2\pi\hbar)^{-1}\int d\tau\,W_{\cal T}(\tau)
\exp(-i{\tau\varepsilon/\hbar})$ is the inverse Fourier transform of
$W_{\cal T}(\tau)$.  The simplest window one can take is the rectangular or
``boxcar" window,
\begin{equation}
  W^{\rm B}_{\cal T}(\tau) = \cases{1 & for $|\tau|\le {\cal T}$, \cr 0 
        & otherwise.}  \qquad
  w^{\rm B}_{\cal T}(\varepsilon) = 2{\cal T}
         \mathop{\rm sinc}\left({\cal T}\varepsilon\over\hbar\right) \;.
  \label{Boxcar}
\end{equation}
$w^{\rm B}_{\cal T}(\varepsilon)$ contains long oscillating tails, which
fall off asymptotically as $\mathop{\rm O{}}(\varepsilon^{-1})$. Since this
falloff is only harmonic it may be possible for oscillations from many
different delta peaks to interfere coherently to yield large spurious peaks.
Thus it may be advisable to choose a different window function.

The problem of estimating the spectrum of an infinite sequence from a finite
subset is a common one in signal processing, and a variety of windows, with
varying properties, have been developed \cite{Harris78}. One of the more
popular is the Hanning window,
\begin{equation}
  W^{\rm H}_{\cal T}(\tau) = {1\over 2}\left[
                  1+\cos\left(\pi\tau\over{\cal T}\right)\right]\;, \qquad
  w^{\rm H}_{\cal T}(\varepsilon) =
     {\pi\sin\left({\cal T}\varepsilon\over\hbar\right) \over
        2\varepsilon\left[\pi^2-\left({\cal T}\varepsilon\over\hbar\right)^2
                  \right]   }\;.
    \label{Hanning}
\end{equation}
The largest sidelobe (i.e.\ tail oscillation) of $w^{\rm H}_{\cal
T}(\varepsilon)$ is $\sim 0.025$ the size of the main lobe, vs.\ $\sim 0.22$
for the rectangular window. Moreover, the tails of $w^{\rm H}_{\cal
T}(\varepsilon)$ decay asymptotically as $\mathop{\rm
O{}}(\varepsilon^{-3})$, and so one need not worry about the cumulative
effect of distant tails.

In order to derive the SCA for $\tilde d_{\cal T}(\varepsilon)$ we start
from the Gutzwiller sum, which approximates the fluctuating part of the level
density by a sum over classical periodic orbits \cite{GutzwillerBook},
\begin{equation}
  \tilde d^{\,\text{sc}}(E) = {2\over\hbar}\sum_p \sum_{r=1}^\infty
         A_{p,r}(E)
         \cos\left({rS_p(E)\over\hbar} - {r\nu_p\pi\over 2}\right)\;.
         \label{Gutzwiller}
\end{equation}
Here the summation is over classical primitive periodic orbits $p$ and
repetitions $r$. $S_p$, $T_p$, and $\nu_p$ are the action, period, and
Maslov index of the primitive orbit $p$, respectively. $A_{p,r}(E) =
T_p/|\det(M_p^r-I)|^{1/2}$ is the stability prefactor, where $M_p$ is the
monodromy matrix for orbit $p$. We also restrict the energy range so that
$E_0$ is large enough to be in the semiclassical domain, while $\Delta E$ is
small enough so that $A_{p,r}(E)$ and $T_p(E)$ can be approximated by their
values at $E_0$, and the action can be replaced by its linearized form,
$S_p(E) \approx S_p(E_0) + \varepsilon T_p(E_0)$. In the following
quantities without an energy argument refer to their values at $E_0$.

Next, we insert (\ref{Gutzwiller}) into (\ref{dtau}) and exchange the order
of the $p$ summation and the $\varepsilon$ integration. The validity of this
step constitutes our primary assumption, and is equivalent to ordering
(\ref{Gutzwiller}) by orbit period.  We can then perform the inverse Fourier
transform on each term of the sum to get
\begin{eqnarray}
  \hat d^{\,\text{sc}}(\tau) &=&  \sum_p \sum_{r=1}^\infty
         A_{p,r}\Biggl\{
          \hat\chi\left(\tau+rT_p\over\hbar\right)
          \exp\left[ir\left({S_p\over\hbar} - {\nu_p\pi\over 2}\right)\right]
       \nonumber\\
           && +
         \hat\chi\left( \tau-rT_p\over\hbar\right)
         \exp\left[-ir\left({S_p\over\hbar} - {\nu_p\pi\over 2}\right)\right]
            \Biggr\}         \;,\label{Fsums} 
\end{eqnarray}
where $\hat\chi(\tau/\hbar)$ is the Fourier transform of the window function
$\chi(\varepsilon)$. For a Gaussian $\chi(\varepsilon)$ it is also a
Gaussian, $\hat\chi(x) = \sqrt{2\pi\sigma^2}\exp[-(\sigma x)^2/2]$.  One can
readily verify that (\ref{Fsums}) is then absolutely convergent, since
$\hat\chi$ effectively selects for each $\tau$ only these orbits (primitive
and otherwise) whose period is within a few $\hbar/\sigma$ of $\tau$.

Finally, inserting (\ref{Fsums}) into (\ref{PartialSum}) and exchanging the
summation and integration gives
\begin{equation}
  \tilde d^{\,\text{sc}}_{\cal T}(\varepsilon) = {2\over\hbar}
  \mathop{\Re {\rm e}}\sum_p\sum_{r=1}^\infty A_{p,r}\,
    {\cal F}_{\cal T}(rT_p,\varepsilon)
      \exp\left[ir\left({S_p+\varepsilon T_p\over\hbar}-{\nu_p\pi\over 2}
         \right)\right]  \label{Finald}\;,
\end{equation}
where ${\cal F}_{\cal T}(T,\varepsilon)$ is a function which depends
on the two windows we used,
\begin{equation}
  {\cal F}_{\cal T}(T,\varepsilon) = {1\over 2\pi}\int_{-\infty}^{\infty} 
    d\tau\; W_{\cal T}(\tau + T)\,  \hat\chi\left(\tau\over\hbar\right)
    \exp\left(-i{\tau\varepsilon\over\hbar}\right)  \;.
\end{equation}
Since $W_{\cal T}(\tau)$ vanishes for $|\tau|>{\cal T}$ and
$\hat\chi(\tau/\hbar)$ falls off rapidly for $|\tau/\hbar|>\sigma^{-1}$,
$\tilde d^{\,\text{sc}}_{\cal T}(\varepsilon)$ only includes orbits whose
period does not extend much beyond ${\cal T}+\hbar/\sigma$. Thus for any
finite ${\cal T}$ one is left with an essentially finite sum, which gives
the SCA for the well defined quantum mechanical quantity $\tilde d_{\cal
T}(\varepsilon)$. In contrast, an orbit-truncated Gutzwiller sum is not in
itself an SCA to anything.

Another way of defining the difference between the two truncation schemes is
the following. Suppose that the action is exactly linear in the energy (this
is the case for e.g.\ the wavenumber level density for billiards). Then,
approximating the whole spectrum using orbit-truncation corresponds to first
taking $\sigma\to\infty$ in (\ref{Finald}), and then ${\cal T}\to\infty$.
However, from (\ref{Finald}) it is clear that one should first take the
${\cal T}\to\infty$ limit to get the local spectrum, and only then extend to
the whole real axis by taking the $\sigma\to\infty$ limit. Since the energy
and time to infinity limits do not in general commute, the two truncation
schemes can give differing results.

We now test the Fourier truncation method for a particular model, the zeros
of the Riemann zeta function on the critical line $z={1\over 2}-iE$. This is
a mathematical, rather than a physical, model.  However, it is well known
that these zeros behave outwardly like the eigenvalues of a chaotic system
without any symmetries (see e.g.\ \cite{KeatingVarenna}). Moreover, the
fluctuating ``level" density for the Riemann zeros on the critical line is
exactly given by
\begin{equation}
  \tilde d(E) = - 2\sum_p\sum_{r=1}^\infty {\log p\over 2\pi\sqrt{p^r}}
     \cos(rE\log p)\;,  \label{ZetaDensity}
\end{equation}
where the $p$ are the prime numbers. Eq.~(\ref{ZetaDensity}) closely
resembles the Gutzwiller trace formula (\ref{Gutzwiller}), with $\hbar=1$,
$\nu_p=0$, $T_p = \log p$, $S_p = E\log p$, and $A_{p,r} = -\log p/
2\pi\sqrt{p^r}$ (note that here the action is exactly linear in the energy).
This ``system" is often used in numerical investigations of the properties
of various spectral sums, since it is a simple matter to generate millions
of prime numbers, while finding periodic orbits in real dynamical systems is
a laborious process at best.

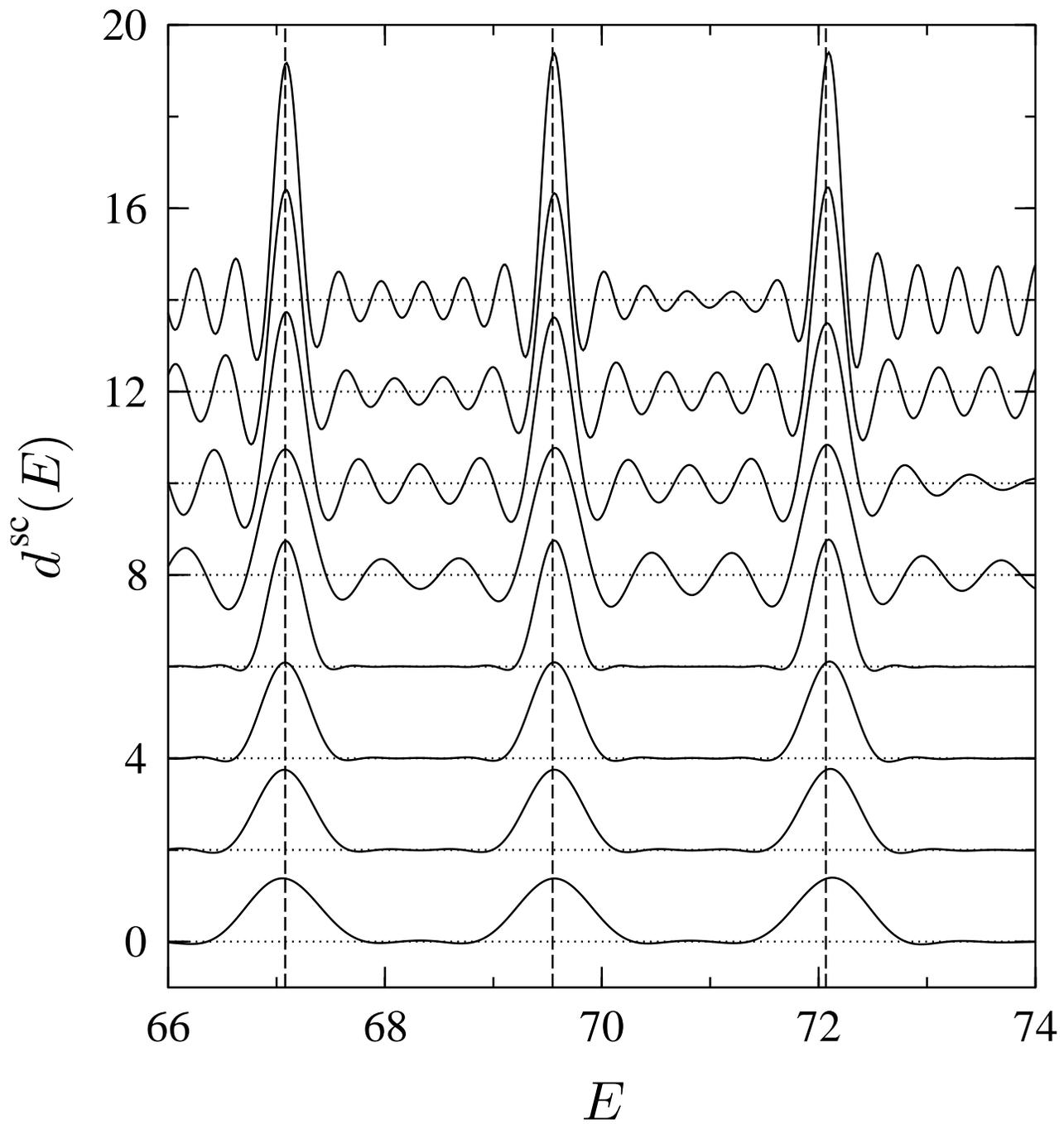
\begin{figure}

\caption{
The level density computed using Fourier truncation, using (from bottom to
top) 1000, 10,000, 100,000 and 5,761,456 primes.  Top 4 lines: Truncating
using a rectangular (``boxcar") window.  Bottom 4 lines: Truncating using a
Hanning window.  Each line is shifted upwards by 2 relative to the preceding
one.  The dashed lines indicate the location of the true Riemann zeros.
}%
\label{Fig:Zeta}
\end{figure}

Using (\ref{ZetaDensity}) the equivalent expressions to (\ref{Fsums}) and
(\ref{Finald}) for the Riemann zeros are
\begin{mathletters}
\begin{eqnarray}
  \hat d(\tau) &=& -\sum_p \sum_{r=1}^\infty
               {\log p\over 2\pi\sqrt{p^r}} \biggl\{
               \hat\chi(\tau+r\log p)\, e^{irE_0\log p}
               +\hat\chi(\tau-r\log p)\, e^{-irE_0\log p} \biggr\}
                \label{RiemannFourier}\\
  \tilde d_{\cal T}(\varepsilon) &=& 
          -2\mathop{\Re {\rm e}}\sum_p\sum_{r=1}^\infty 
            {\log p\over 2\pi\sqrt{p^r}} \,
       {\cal F}_{\cal T}(r\log p,\,\varepsilon)\, e^{irE\log p}
       \label{ZetaFunctionSum} \;.
\end{eqnarray}
\end{mathletters}
We test these expressions numerically in Fig.~\ref{Fig:Zeta}. We used 1000,
10,000, 100,000 and 5,761,456 primes (i.e.\ all primes below $10^8$), taking
$E_0=69.5$ (for historical reasons) and a Gaussian $\chi(\varepsilon)$ with
$\sigma=2.5$. ${\cal T}$ was chosen in each case so that all the Fourier
elements considered could be evaluated to a good accuracy using the
available primes (more about this later).

The top 4 lines in Fig.~\ref{Fig:Zeta} show the resulting approximate level
density using a rectangular window $W^{\rm B}_{\cal T}(\tau)$.  We clearly
see the $\delta$ peaks build up as ${\cal T}$ increases. As expected from
(\ref{ApproxDeltas}), the amplitude of the intervening oscillations is not
strongly dependent on ${\cal T}$, but is more or less fixed by the form of
$W_{\cal T}(\tau)$. This is also evident in the bottom 4 lines, where we
show the approximate level density using $W^{\rm H}_{\cal T}(\tau)$. Here
the oscillations are much smaller, at the price of an increase in the width
of the peaks, as expected. We conclude that Fourier truncation appears to
work very well for the Riemann zeros.

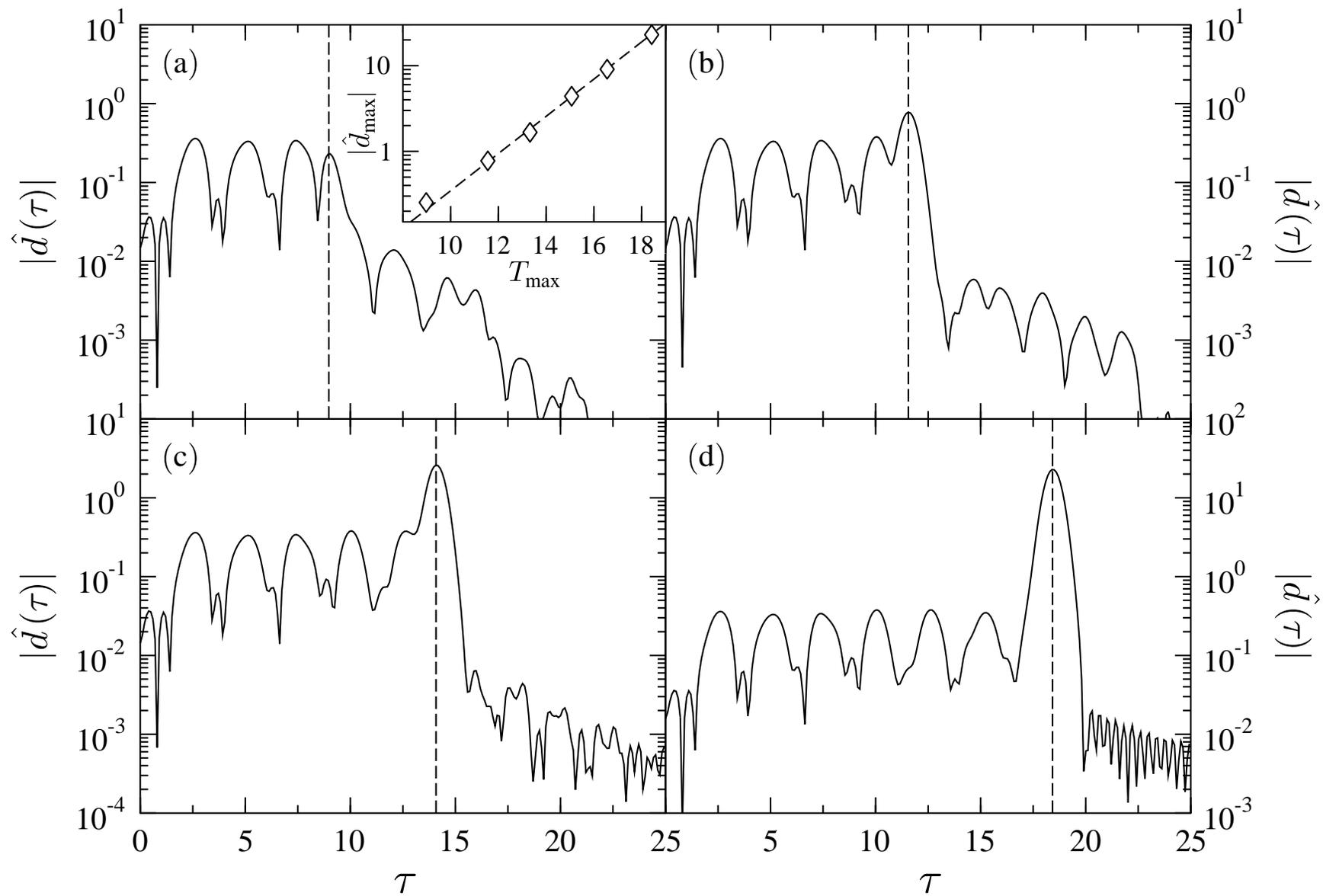
\begin{figure}

\caption{
The Fourier spectrum of the orbit-truncated sum (\protect\ref{ZetaDensity}),
using 1000, 10,000, 100,000 and 5,761,456 primes in subfigures a--d,
respectively. The point $\tau=T_{\text{max}}$ is marked by a vertical dashed
line.  Inset --- $|\hat d_{T_{\text{max}}}(T_{\text{max}})|$ vs.\
$T_{\text{max}}$. The dashed line is the prediction of
(\protect\ref{ExpIncrease}).
}%
\label{Fig:ZetaTrunc}

\end{figure}

It is instructive to compare Fig.~\ref{Fig:Zeta} to the level density
obtained by orbit truncation at the same place. In this case, as more primes
are added to the sum peaks start to emerge, but are eventually obscured by
increasing oscillations \cite{KeatingVarenna}. We can now understand this
behavior by examining Fig.~\ref{Fig:ZetaTrunc}, where we show the Fourier
spectrum of the truncated sum (\ref{ZetaDensity}) using the same primes as
in Fig.~\ref{Fig:Zeta}. It is apparent that, as more primes are added, a
spectral peak builds up near $\tau\approx T_{\text{max}}=\log
p_{\text{max}}$ (denoted by a dashed vertical line), where $p_{\text{max}}$
is the largest available prime. This peak eventually dominates the resulting
level density.

We can understand this peak in the following manner. For large positive
$\tau=T_{\text{max}}$ we can approximate (\ref{RiemannFourier}) by its
second term, and also neglect all repetitions. We can then write $\hat
d_{T_{\text{max}}}(T_{\text{max}})$ in the presence of orbit truncation in
the form of another Fourier transform between the variables $x$ and ${\cal
E}$, taken at ${\cal E}=E_0$,
\begin{eqnarray}
  \hat d_{T_{\text{max}}}(T_{\text{max}},E_0) &\approx& -
      {e^{-iE_0T_{\text{max}}}\over 2\pi} \int dx\;
      \mu(x+T_{\text{max}})\,\hat\chi(x)\,\Theta(-x)
      e^{-i{\cal E}x}\;\Bigl|_{{\cal E}=E_0} \nonumber \\
    &=& \left[ \hat d(T_{\text{max}},{\cal E}) \otimes \hat\Theta^*({\cal E})
           \right]_{{\cal E}=E_0}
       \;,  \label{Fourier2}
\end{eqnarray}
where $\Theta(x)$ is the step function, $\hat\Theta({\cal E})={1\over
2}\delta({\cal E})-i/2\pi{\cal E}$ is its Fourier transform, $\mu(x)$ is the
weighted period density $\mu(x) = \sum_p x\,e^{-{1\over 2} x}\,\delta(x-\log
p)$, and $\otimes$ denotes a convolution between quantities which are
considered functions of ${\cal E}$. Since $\hat\Theta({\cal E})$ contains
long harmonic tails, $\hat d_{T_{\text{max}}}(T_{\text{max}})$ will also
include $(2\pi E_0)^{-1}$ times the DC (${\cal E}=0$) component of $\hat
d(T_{\text{max}})$. From the prime number theorem $\mu(x)\sim\exp(x/2)$, and
so for large enough $T_{\text{max}}$ we get
\begin{equation}
  \left|\hat d_{T_{\text{max}}}(T_{\text{max}})\right| \approx
  {1\over (2\pi)^2 E_0} \int dx\;\mu(x+T_{\text{max}})\,\hat\chi(x)
  \approx {e^{1/8\sigma^2}\over 2\pi E_0}\,
    \exp\left(T_{\text{max}}\over 2\right)\;.
  \label{ExpIncrease}
\end{equation}
Thus the discontinuous truncation ``folds" into $\hat
d_{T_{\text{max}}}(T_{\text{max}})$ an exponentially increasing contribution
from the DC component, giving rise to the peak in Fig.~\ref{Fig:ZetaTrunc}.
The inset of Fig.~\ref{Fig:ZetaTrunc}a shows $\hat
d_{T_{\text{max}}}(T_{\text{max}})$ vs.\ (\ref{ExpIncrease}), and we see
that the above prediction works quite well.

Finally, we note that unlike in the Riemann case, typical dynamical systems
also include Maslov phases. This changes the picture completely, as the DC
term which is responsible for the divergent oscillations is then strongly
suppressed.  We therefore expect that the errors induced by the orbit
truncation procedure would in such cases be less severe.

In this work we proposed writing the SCA to the level density of a
classically chaotic system by means of a limiting process, consisting at
each stage of a convergent SCA to a well defined spectral quantity which
itself converges to the level density in the limit. Numerical tests on the
application of this method to the Riemann zeros yielded increasingly good
approximations to the level density with no divergent oscillations.
Moreover, we were able to understand the oscillations that appear in the
conventional truncation approach as an edge effect resulting from the abrupt
nature of the truncation.


This work was funded by a grant from the MINERVA.


\begin{thebibliography}{1}

\bibitem{Gutzwiller88}
M.~C. Gutzwiller, J. Phys. Chem {\bf 92},  3154  (1988).

\bibitem{KeatingVarenna}
J.~P. Keating,  in {\em Proceedings of the 1991 Enrico Fermi International
  School on ``Quantum Chaos'', Course CXIX}, edited by G. Casati, I. Guarneri,
  and U. Smilansky (Elsevier B.V., Amsterdam, 1993), pp.\ 147--186.

\bibitem{MartinPhD}
M. Sieber, The Hyperbola Billiard: A Model for the Semiclassical Quantization
  of Chaotic Systems, 1991, (PhD Thesis).

\bibitem{Guinand48}
A.~P. Guinand, Proc. Lond. Math. Soc. {\bf 50},  107  (1948).

\bibitem{BerrySchool}
M.~V. Berry,  in {\em Proceedings of the 1989 Les Houches Summer School on
  ``Chaos and Quantum Physics''}, edited by M.-J. Giannoni, A. Voros, and J.
  Zinn-Justin (Elsevier Science Publishers B.V., Amsterdam, 1991), p.\ 251.

\bibitem{Aurich92}
R. Aurich and F. Steiner, Proc. Roy. Soc. Lond. A {\bf 437},  693  (1992).

\bibitem{Harris78}
F.~J. Harris, Proc. IEEE {\bf 66},  51  (1978).

\bibitem{GutzwillerBook}
M.~C. Gutzwiller, {\em Chaos in Classical and Quantum Mechanics}
  (Springer--Verlag, New York, 1990).

\end{thebibliography}

\newpage
\widetext

\end{document}